\documentstyle[12pt,epsbox]{article}

\begin{document}

\pagenumbering{arabic}

\begin{center}
{\bf Long periodicity of Blazar with RXTE ASM, TA and HEGRA}

{Satoko Osone$^a$ and Masahiro Teshima$^b$}

$^a${\it Matsudo 1083-6 A-202, Matsudo, Chiba 271-0092, Japan}

$^b${\it Institute for Cosmic Ray Research, University of Tokyo, 5-1-5 Kashiwa-no-Ha, Kashiwa City,Chiba 277-8582, Japan} 

{osone@icrr.u-tokyo.ac.jp}

\end{center}

\section*{Abstract}
Long periodicity for Mkn501 during a large flare in 1997 have been reported by TA, HEGRA group.
Here, we establish this periodicity with archival data of RXTE All Sky Monitor(ASM), Telescope Array(TA) and HEGRA with a chance probability less than $10^{-5}$.
We also find that an origin of 23 day periodicity is related with a change of either a gamma factor of electrons $\gamma$ or the magnetic field or a beaming factor.
And, in order to search for a category which have a long periodicity, we make a periodicity analysis for ten H.B.L. with RXTE ASM. We find a long periodicity for three TeV detected source, Mkn501, Mkn421 and PKS2155-304.
There may exist a relation between an origin of a long periodicity and an electron acceleration to a TeV range.

\section{Introduction}
TeV gamma ray flux of HBL Mkn421 and Mkn501 are usually lower than that of Crab.
However, in 1997, there was a large flare of Mkn501, the recorded flux grew to ten times  that of Crab(Aharonian et al, 1999a, 1999b, 1999c, Krawczynski et al.2000).
During this time, a 13 day periodicity in TeV gamma ray flux of Mkn501 was reported initially by the TA (Hayashida et al.1998). A periodicity of 23 day was reported by the TeV cherenkov detectors HEGRA, Telescope Array and by the Xray detector {\it RXTE} ASM (Kranich et al. 1999, Nishikawa et al.1999).
However, a 23 day periodicity for Mkn501 in 1997 has not been established yet.
In this paper, we reanalyze both the RXTE ASM data and the TeV data of TA and HEGRA.
In order to search for  the kind of AGN which have a long periodicity,
we do a periodicity analysis for ten HBLs using the archive data of the Xray satellite {\it RXTE} ASM.
We also search for an energy spectral variation with a possible periodicity with RXTE PCA.

\section{Observation}
We obtain data points of (MJD, rate, error) for Mkn501 in 1997 with TA(Hayashida et al. 1998) and with HEGRA(Kranich 2001).
We obtain RXTE ASM data of 90 sec integrated data of (MJD,rate,error) from Jan. 1996(MJD 50087) to Aug. 2000(MJD 51780).
We collect ten HBLs cataloged by Xie et al.(1993) from ASM archives. 
We obtain an archive data with RXTE PCA for a spectral analysis.
MJD 50300-50900 is used for a confirmation of 23 day periodicity with ASM data.
For ASM data, because of poor statistics, we make several kind of data set for each target with both a binning and a selection of a ratio of a rate and an error.
We use Lomb-Scargle method(Lomb 1976; Scargle 1982), which is suitable for unevenly sampled data.
We calculate a chance probability with Prob($\ge z$) $\equiv$ $ 1 - (1 - e^{-z} )^N$($z$ is a power, $N$ is a number of data).
For ASM data, we limited the search interval for the periodicity between 1 day and $T_{obs}$/10, in order to obtain more than 10 cycles for a periodicity. Here, $T_{obs}$ is the total observation time.
 
\section{periodicity of Mkn501 in 1997}
We show results at table 1, figure 1.
We confirmed 22-23 days periodicity for Mkn501 during a flare in 1997 with both TeV gamma ray data of TA, HEGRA and RXTE ASM data.
22-23 days periodicities observed by RXTE ASM and TA and HEGRA are same within FWHM.

There is a relation of power$\sim f^{-1}$ ($f$ is a frequency) in an Xray binned power spectra, which is popular in AGN as figure 2.
This trend may have a physical meaning, though this has not yet been resolved.
The periodicity we detect may be a fluctuation of this trend.
For TeV gamma ray binned power spectra, there does not seems to be a $f^{-1}$ component as figure 2.
Even if there may be also a $f^{-1}$ component which has a physical meaning in TeV gamma ray power spectra, we obtain a chance probability less than $10^{-5}$ for 23 days periodicity of Mkn501 in 1997 by combining results of  ASM data and TA data and HEGRA data.
We show the summed power spectra of Xray data of RXTE ASM and TeV gamma ray data of TA and HEGRA as figure 3. 
Each power is normalized with $f^{-1}$ component.

\begin{table}
\caption{Summary of periodicity analysis. A chance probability$^*$ is for the power spectra which has
 a $f^{-\alpha}$ component. The error of periodicity is FHWM. The two different periodicity for Mkn501 is attributed to a different kind of data set.}
\begin{center}
\begin{tabular}{ccrcc}\hline
target & data & period(day) & chance & chance\\
     &   &   &   probability &  probability$^*$ \\ \hline
Mkn501 & RXTE ASM & 152.5 $+7.1 -7.3$ & 3$\times10^{-4}$   & 2.3$\times10^{-1}$ \\
       &          & 54.4 $+0.9 -0.8$  & 1.9$\times10^{-4}$   & 1.9$\times10^{-1}$ \\ \hline
Mkn501 in 1997 &  RXTE ASM & 23.7 $+0.5 - 0.4$ & 1.7$\times10^{-13}$ & 7.7$\times10^{-2}$  \\
               & TA   & 23.7 $+ 2.1 - 1.7$ & 1.4$\times10^{-2}$    & 2.8$\times10^{-1}$ \\
               &  HEGRA & 22.5 $+1.9 -1.7$& 3.4$\times10^{-11}$ & $\le10^{-5}$   \\ \hline
Mkn421 & RXTE ASM  & 102.5 $+3.9 - 3.4$ & 6.7$\times10^{-6}$ &  1.8$\times10^{-2}$\\ \hline
PKS2155-304 & RXTE ASM & 142.5 $+5.5 -5.5$ & 2.2$\times10^{-8}$ & $\le10^{-3}$   \\ \hline
\end{tabular}
\end{center}
\end{table}

\begin{figure}[t]
  \begin{center}
    \psbox[height=4cm]{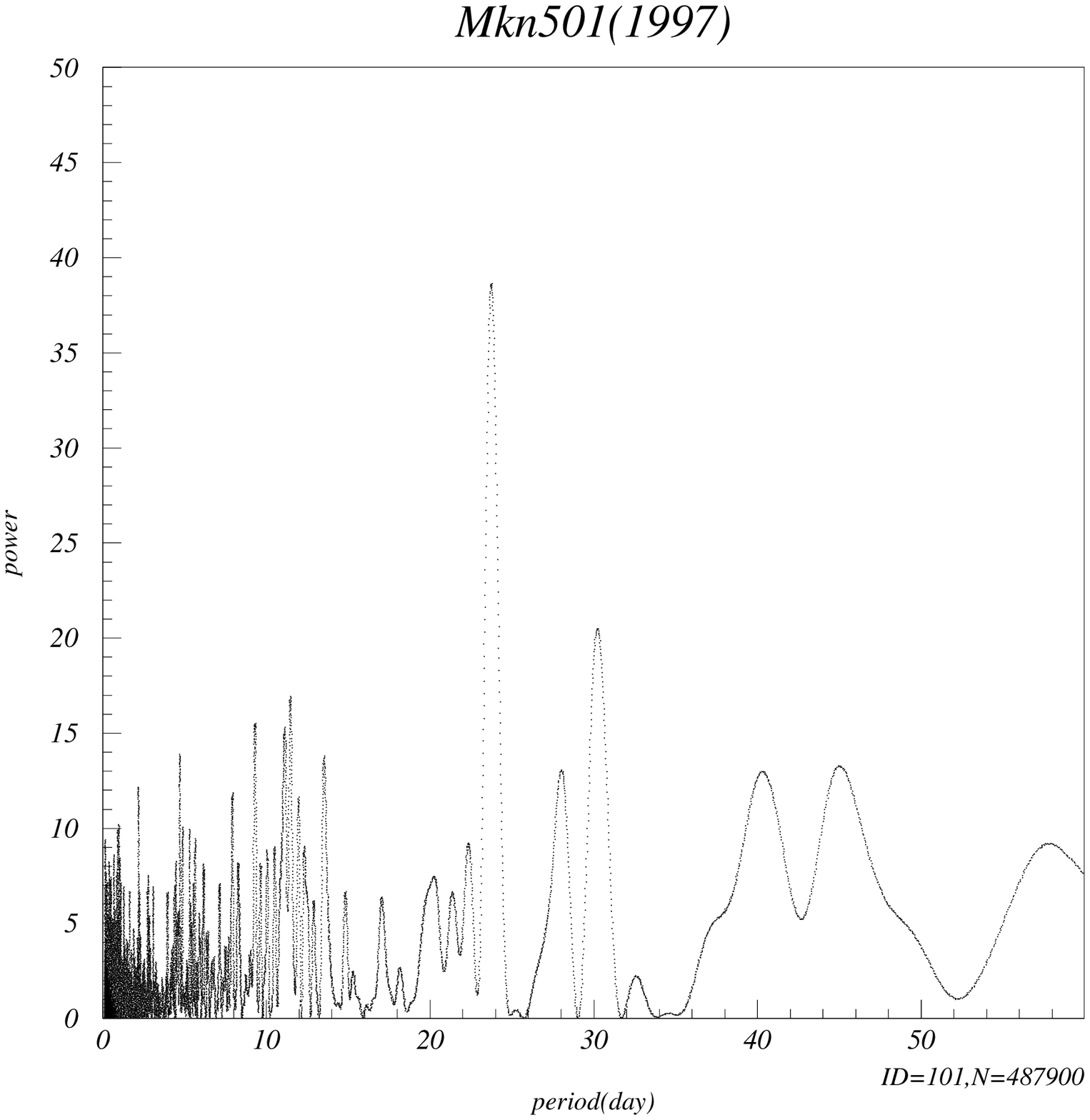}
    \psbox[height=4cm]{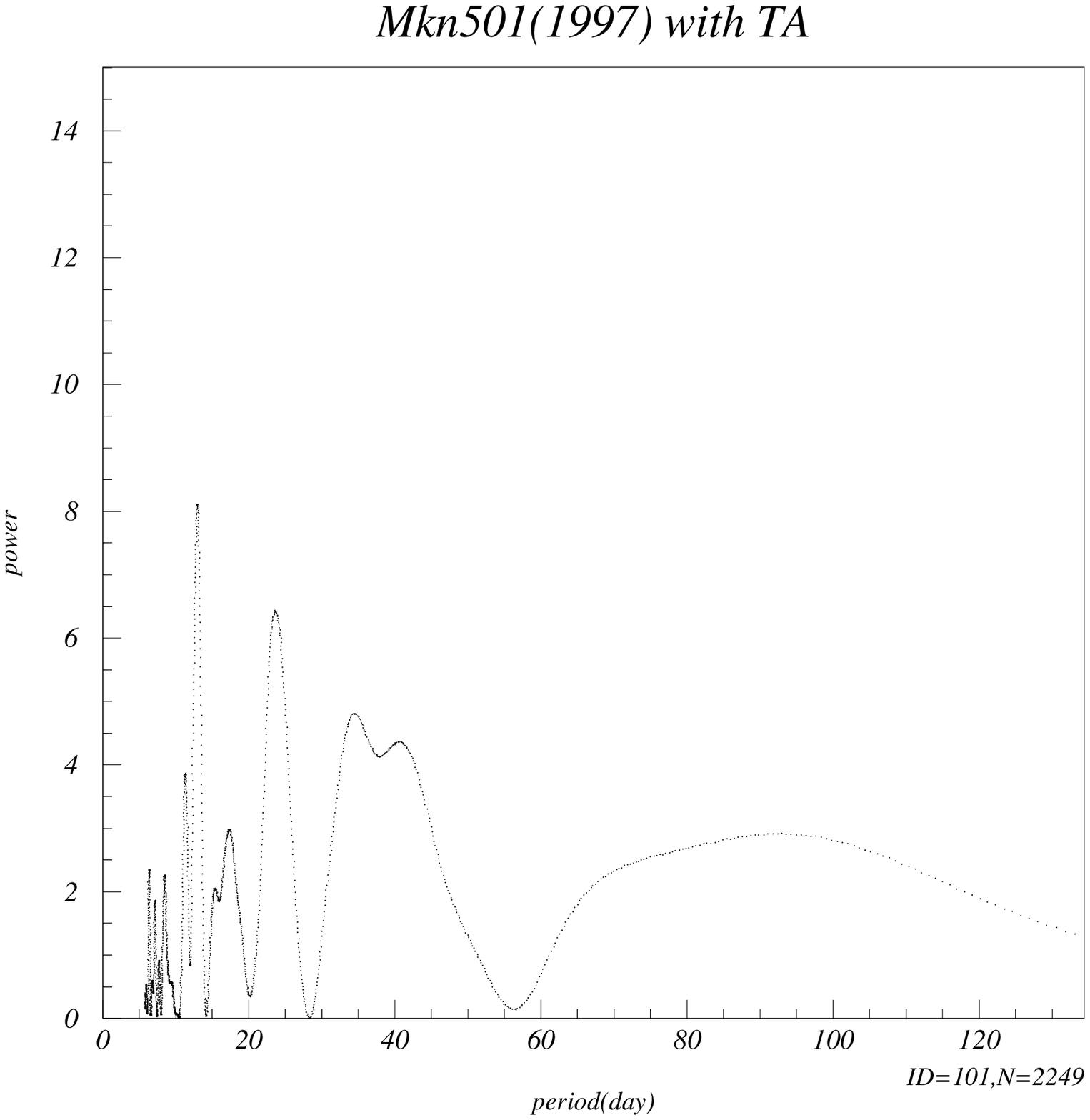}
    \psbox[height=4cm]{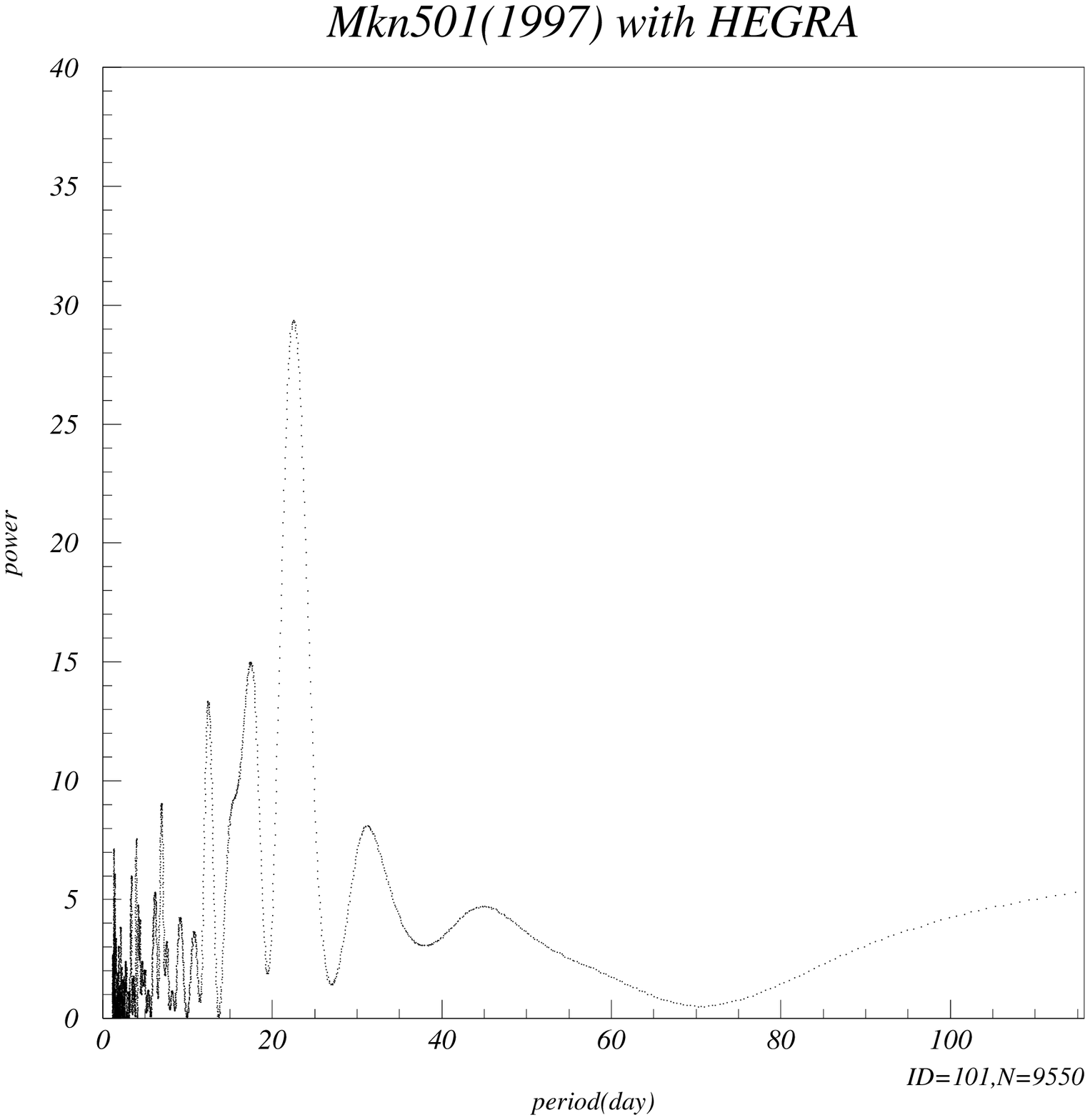}
  \end{center}
  \caption{The Xray power spectra with RXTE ASM(left) and TeV gamma ray power spectra with Telescope Array(middle) and HEGRA(right) for Mkn501 in 1997.}
\end{figure}

\begin{figure}[t]
  \begin{center}
    \psbox[height=4cm]{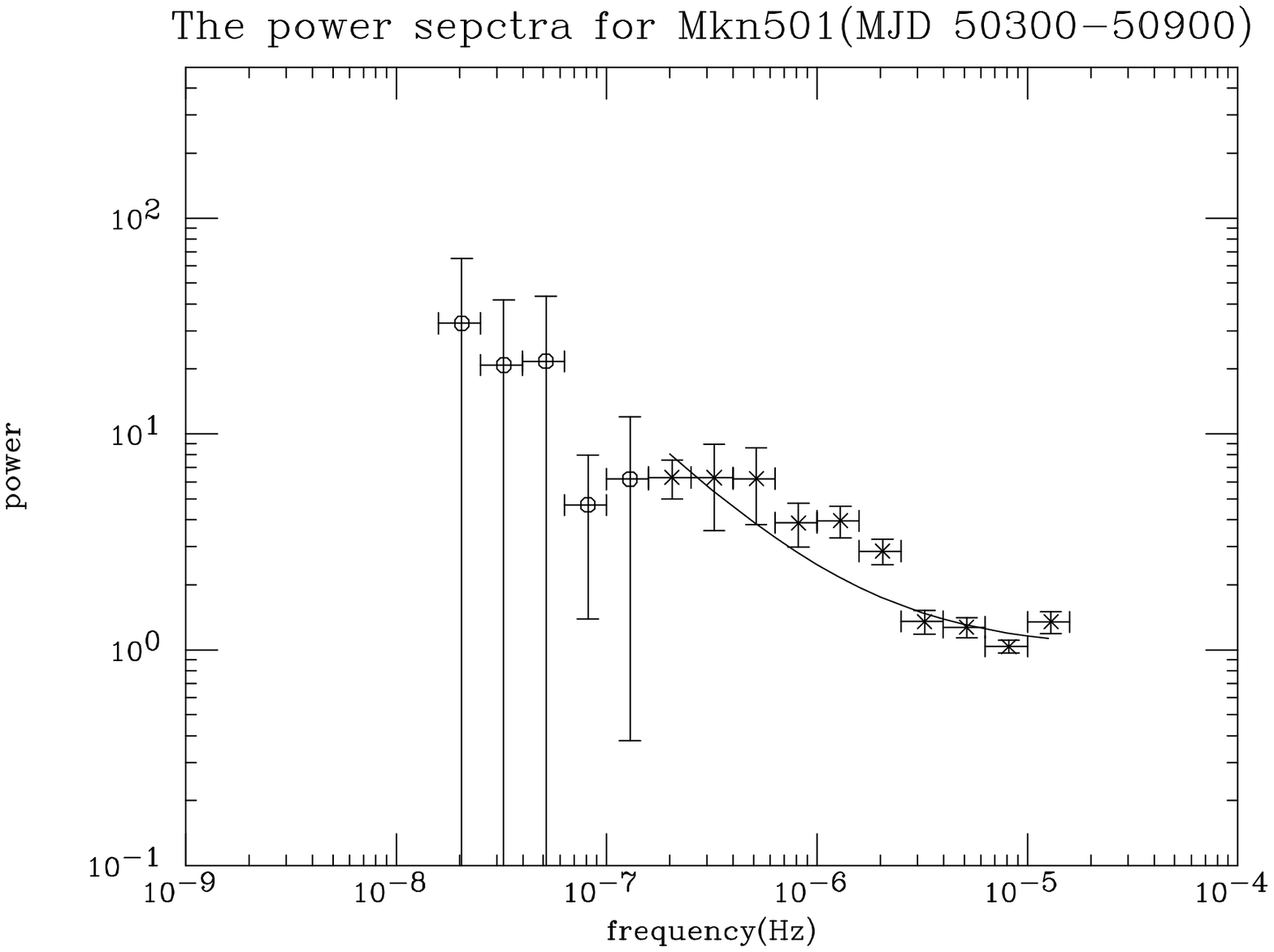}
    \psbox[height=4cm]{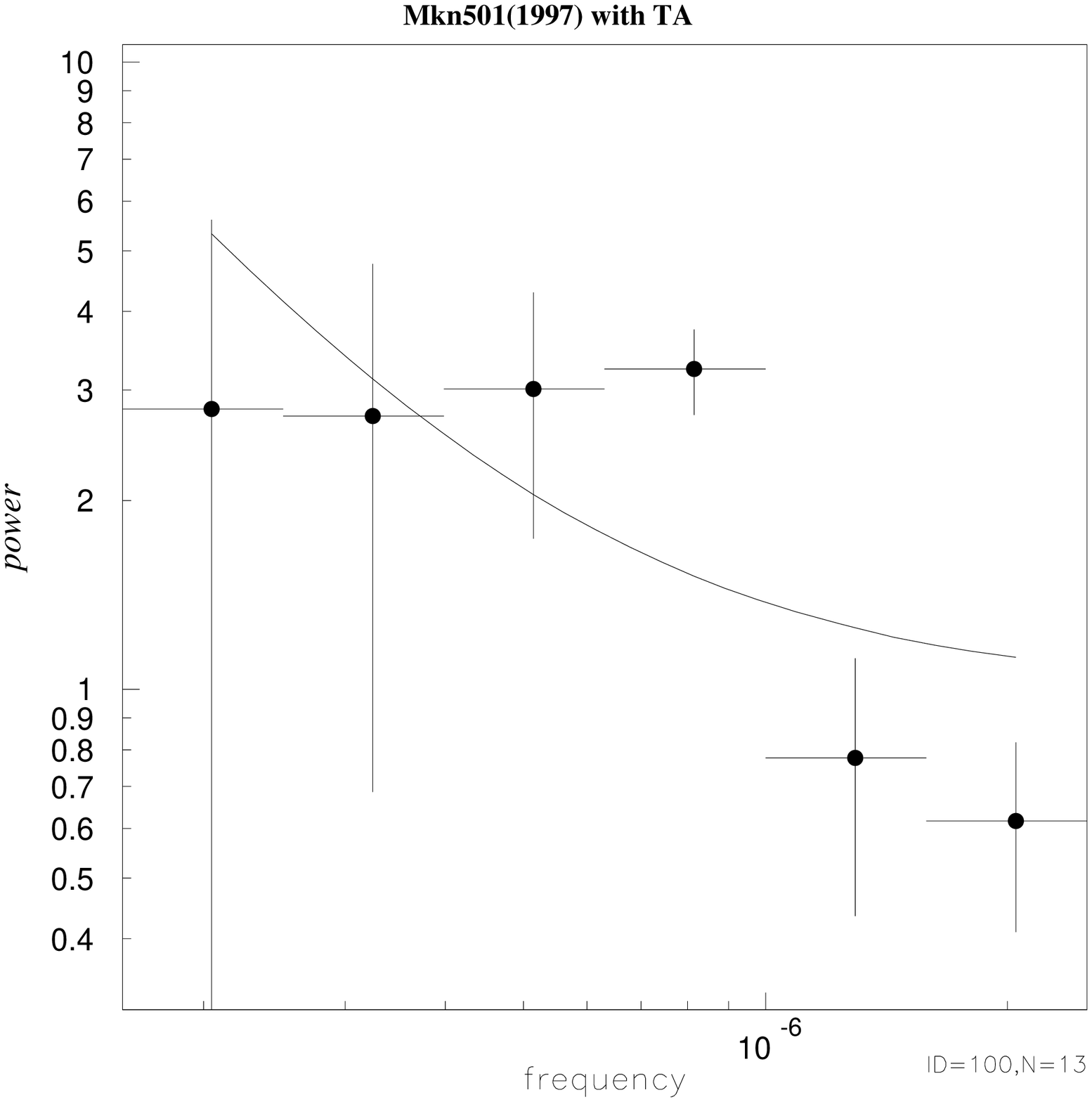}
     \psbox[height=4cm]{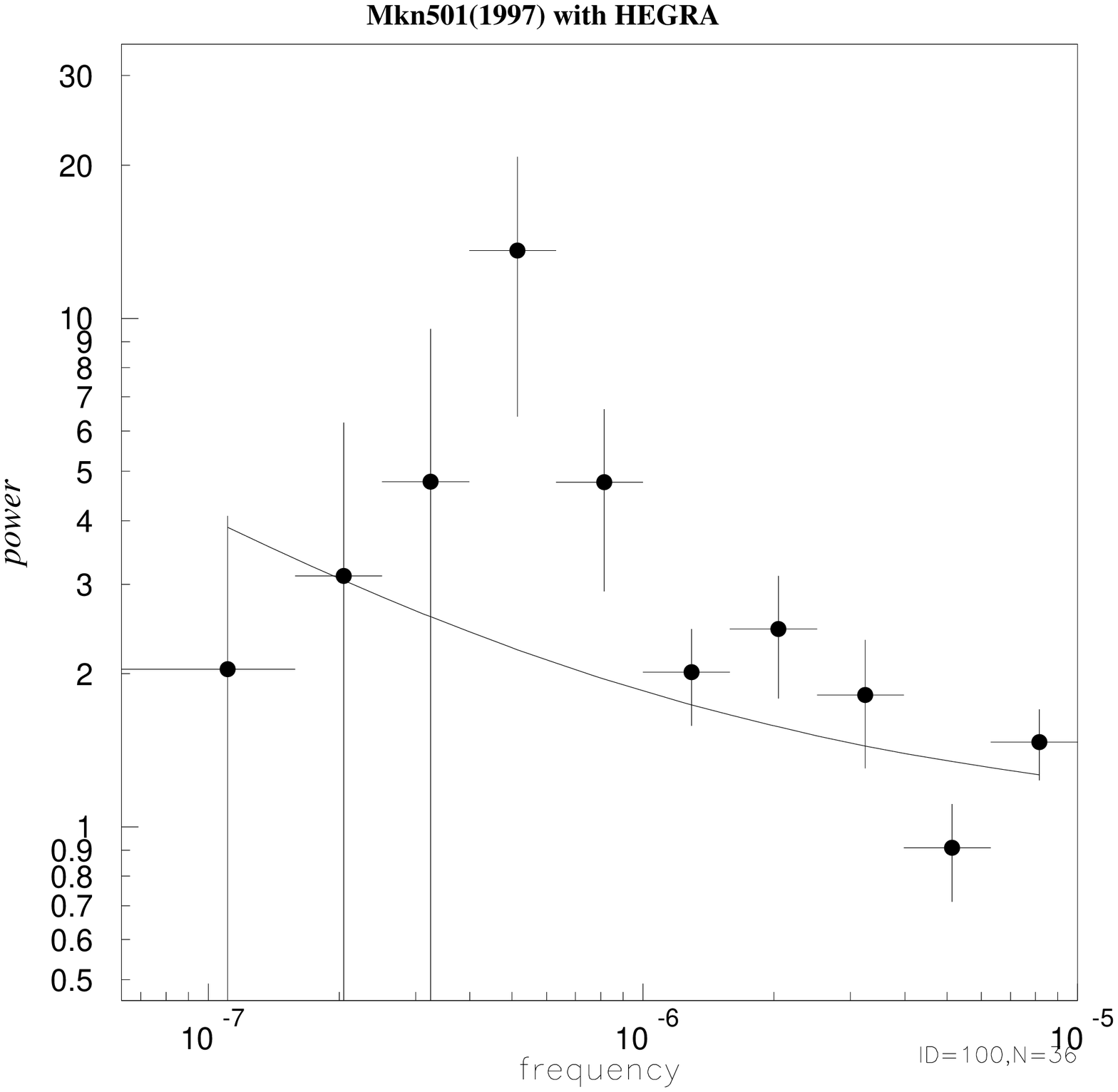}
  \end{center}
  \caption{The Xray binned power spectra with RXTE ASM(left) and TeV gamma ray binned power spectra with Telescope Array(middle) and HEGRA(right) for Mkn501 in 1997. A solid line is best fitted model of $1+A*f^{-\alpha}$($A$ is a constant and $f$ is a frequency).}
\end{figure}

\begin{figure}[t]
  \begin{center}
    \psbox[height=4cm]{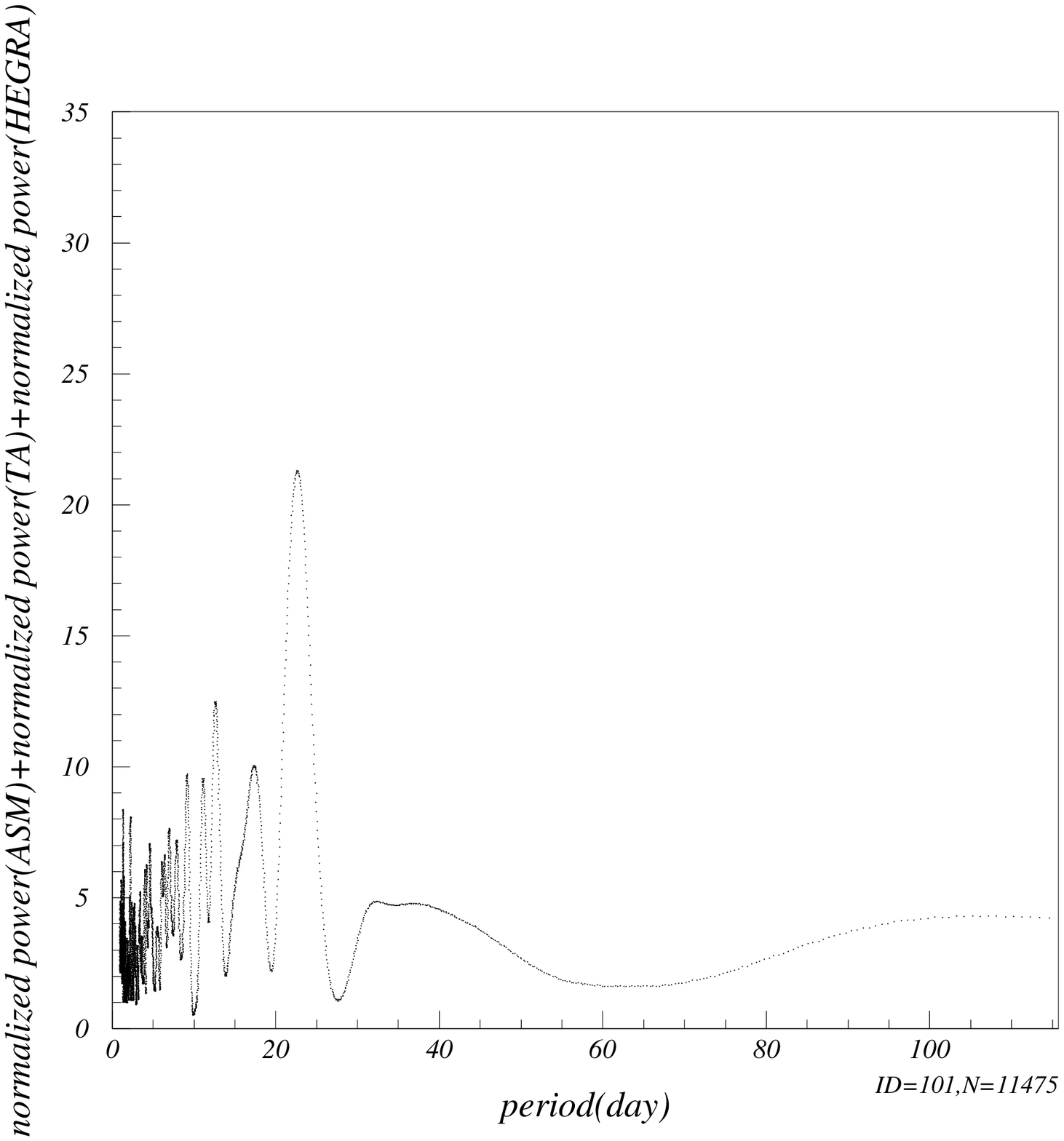}
  \end{center}
  \caption{The summed power spectra of Xray data of RXTE ASM and TeV gamma ray data of TA and HEGRA.  Each power is normalized with $f^{-\alpha}$ component.}
\end{figure}

\section{Origin of 23 day periodicity for Mkn501 in 1997}
We show a phase diagram of a photon index of an Xray energy spectra with RXTE PCA for Mkn501 in 1997 at figure 4.
There is a variation of a photon index during one period.
In an energy spectra of $\nu F_{\nu}$(erg s$^{-1}$)vs.$\nu(Hz)$, an index of a spectra becomes $-a +2$(Here, $a$ is a photon index). The change of a photon index from 1.8 to 2.4 implies an index in this spectra from 0.2 to $-$0.4.
This suggest that the peak energy of a synchrotron radiation is changing over the PCA energy range during a flare time.
This argument is consistent with Kataoka (2000).
We find a periodic change of a peak energy of a synchrotron radiation during one period.
The peak energy  for a synchrotron radiation is dependent on a gamma factor of electrons $\gamma$, the magnetic field and a beaming factor (Rybicki \& Lightman 1979).
Therefore, we expect a periodic change of these parameters, which related with origins of the periodicity, during a flare for Mkn501.

\begin{figure}[t]
  \begin{center}
    \psbox[height=4cm]{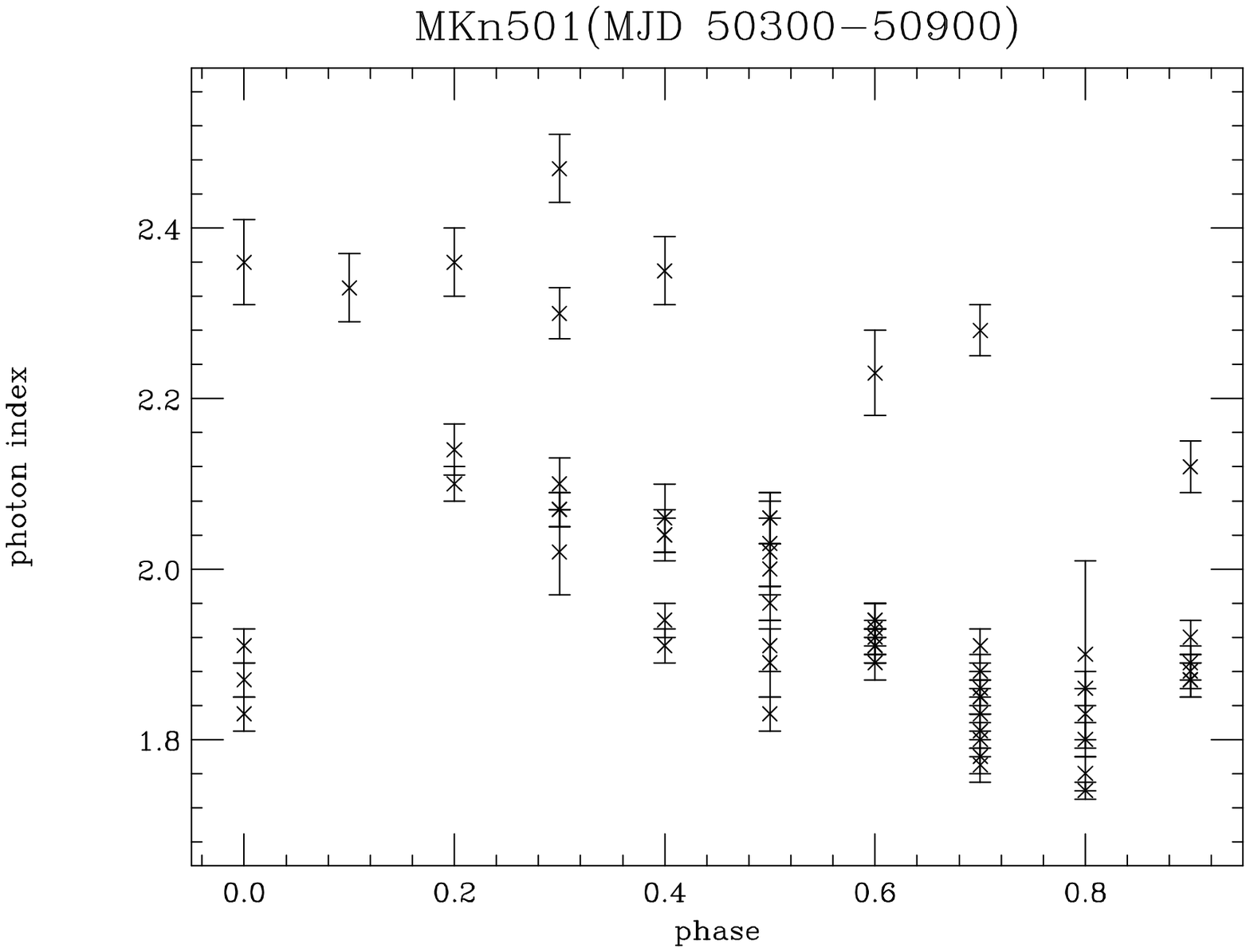}
  \end{center}
  \caption{a phase diagram of a photon index of an Xray energy spectra with RXTE PCA for Mkn501 in 1997.}
\end{figure}

\section{Category which have a long periodicity}
We detect possible periodicities of 10-100day for three Blazars, Mkn421, Mkn501 and PKS2155-304 among ten Blazars. We show these power spectra at figure 5. We show results at table 1.
The periodicities detected for three Blazars emit TeV gamma ray emission have been detected (Mkn 421;eg.Aharonian et al. 1999b, Mkn 501;eg. Aharonian et al. 1999d, PKS 2155-304;P.M.Chadwick et al. 1999).
Then, the probability that all periodicity detected Blazars are all TeV gamma ray detected one is  8.3$\times10^{-3}$.
There may exist a relation between an origin of a long periodicity and an electron acceleration to a TeV range.
However, these periodicities are less significant when $f^{-1}$ component
has a physical meaning as table 1.

\begin{figure}[t]
  \begin{center}
    \psbox[height=4cm]{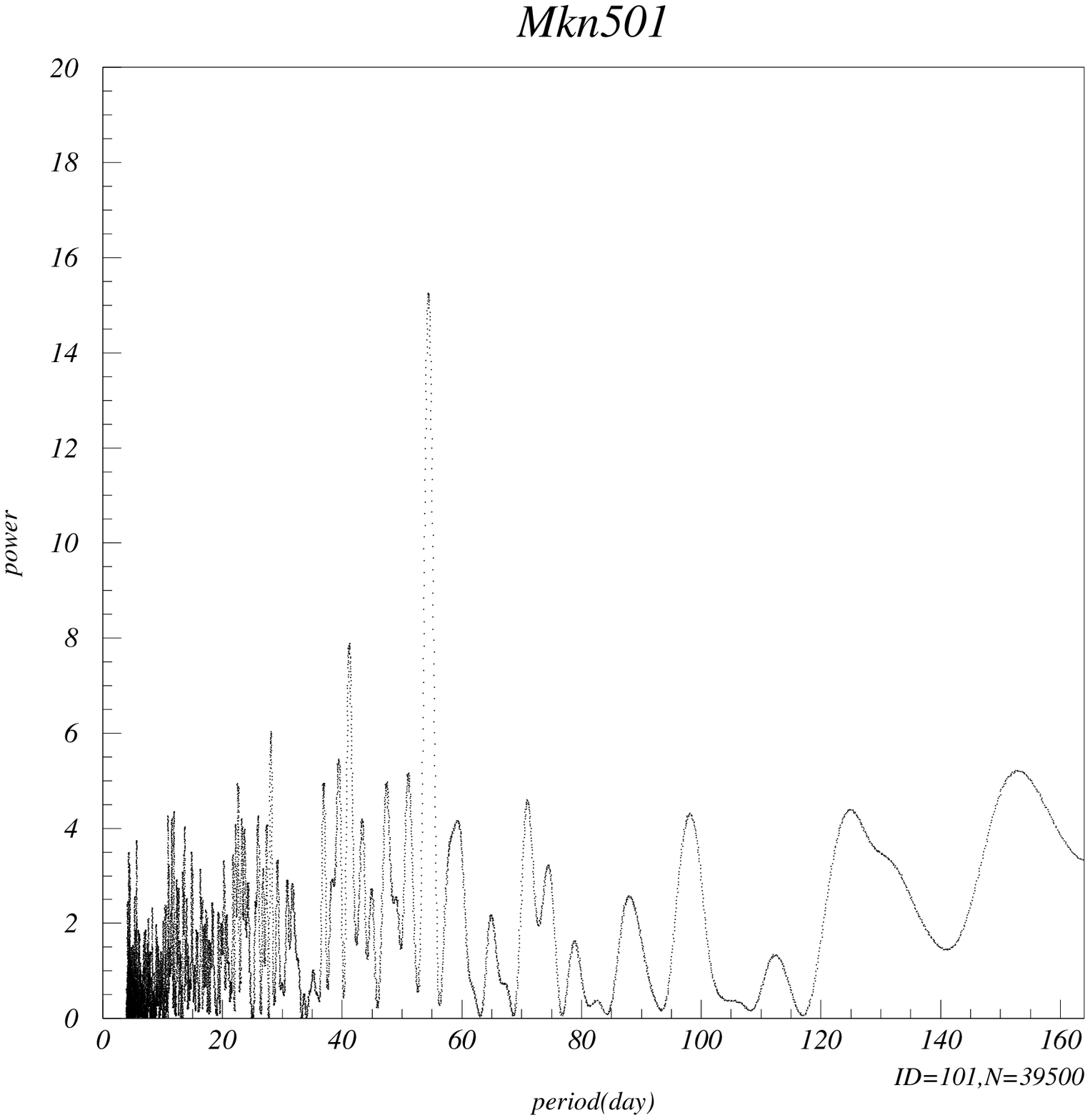}
    \psbox[height=4cm]{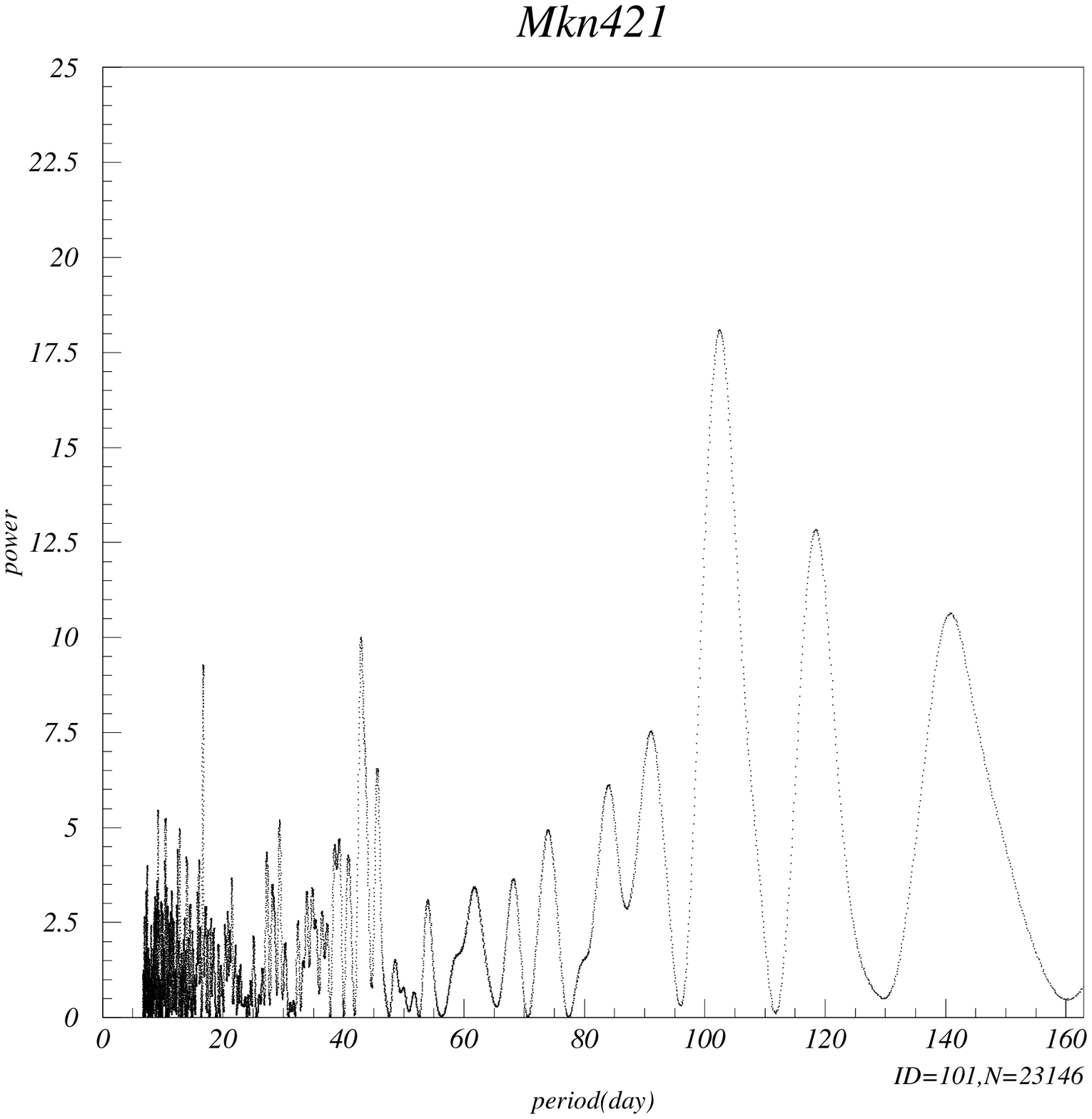}
    \psbox[height=4cm]{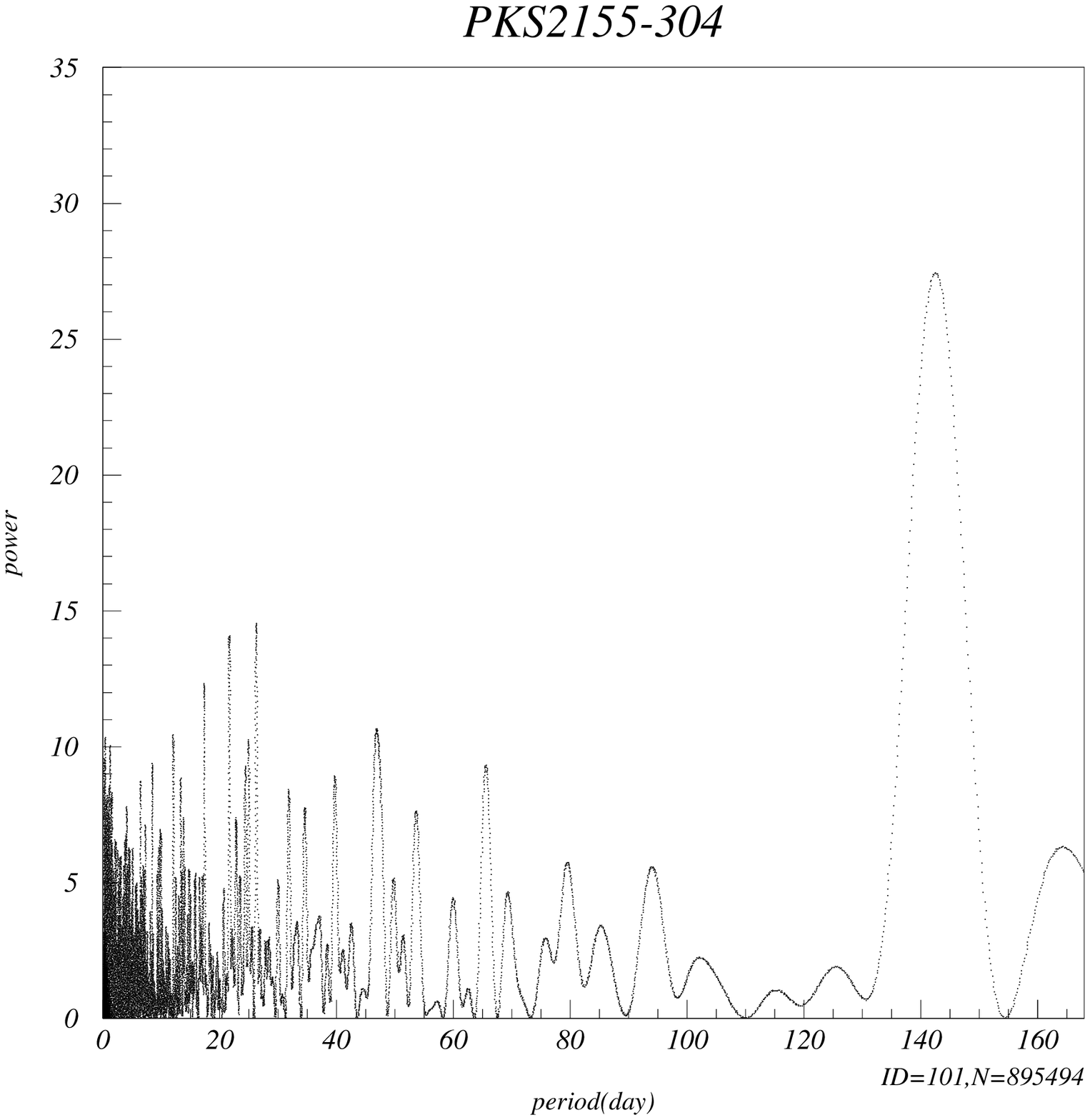}
  \end{center}
  \caption{The Xray power spectra with RXTE ASM for Mkn501(right), Mkn421(middle) and PKS2155-304(right) for 4.6 years.}
\end{figure}

\section{References}

\begin{itemize}
\setlength{\itemsep}{-1.5mm}
\setlength{\itemindent}{-8mm}
\item[]1.Aharonian F.A. et al, A\&A, 349,29,1999a
\item[]2.Aharonian F.A. et al, A\&A, 350, 757, 1999b
\item[]3.Aharonian F.A. et al, A\&A, 349,11,1999c
\item[]4.Aharonian F.A. et al., A\&A, 342,69,1999d
\item[]5.Chadwick P.M. et al., ApJ, 513, 161, 1999.
\item[]6.Hayashida H. et al., ApJ, 504, L71, 1998.
\item[]7.Kataoka Doctor thesis, University of Tokyo, 2000.
\item[]8.Kranich D. et al., Proceeding 26$^{th}$ICRC, OG 2.1.18, 1999.
\item[]9.Kranich D., Doctor thesis,``Temporal and spectral characteristics of the AGN Mkn 501 during a phase of high activity in the TeV range'', 05.2001,  MPI Munich
\item[]10.Krawczynski H. et al, A\&A, 353,97,2000
\item[]11.Lomb N.R., Astro. and Spa. Sci, 39,447, 1976.
\item[]12.Nishikawa D.et al.,  Proceeding 26$^{th}$ICRC, OG 2.1.17, 1999.
\item[]13.Rybicki G.B. \& Lightman A.P., 1979, {\it Radiative Processes in Astrophysics}(John wiley \& Sons)
\item[]14.Scargle  J.D., ApJ, 263, 835, 1982.
\item[]15.Xie G.Z. et al., A\&A, 278, 6, 1993.
\end{itemize}

\end{document}